\documentstyle[aps,prb,epsf,twocolumn]{revtex}

\begin{document}
\draft
\twocolumn[\hsize\textwidth\columnwidth\hsize\csname @twocolumnfalse\endcsname

\title{A numerical study of the Random Dimerized XX spin-1/2 chain}
\author{P. Henelius and S.M. Girvin}
\address{Department of Physics,
Indiana University, Bloomington, IN 47405}
\date{\today}  
\maketitle

\begin{abstract}
The effects of randomness and dimerization on the spin-1/2 XX chain
are studied by a mapping to free fermions. Results are presented
for the transverse and longitudinal components of the average and
typical spin and string correlation functions.
Contrary to previous numerical evidence, the decay exponents of all 
the above correlation functions are found to be in good agreement 
with the theoretical RG predictions for the random singlet phase.
We also present cross-over functions for the above correlation functions
in the random dimer phase. Disorder averages have been
taken for system sizes up to N=1024, about ten times larger
than in previous studies.
\end{abstract}
\vskip2mm]

\section{INTRODUCTION}
It is the purpose of this work to study the effects
of disorder and dimerization on the ground state of the 
XX spin-1/2 chain. Much analytical
\cite{fish1,fish2,fish3,hyma,hyma2,west1,west2,bale,kenz} and 
numerical\cite{youn,rode,hida,hida2,haas} work has been done on random spin 
chains, as well as higher dimensional spin models\cite{guom,rieg,sent}
and to begin with we will review some of the knowledge we have of these 
systems.

In 1983 Haldane pointed out\cite{hald} that integer spin chains have an 
energy gap, while half integer spin chains are gapless. If we introduce 
alternating bond strength in a spin-1/2 chain, and thereby enforce
dimerization, an energy gap is induced\cite{sing}. In the absence of 
disorder the spin-1 and dimerized spin-1/2 chains are in the same 
phase\cite{hida}, since one can be continuously transformed into the other. 
This Haldane phase is characterized by an excitation gap, exponential decay of
the spin correlation function and a non-vanishing string order parameter,
to be defined below. 

Fisher\cite{fish1,fish2,fish3} has recently made considerable progress 
towards understanding
the strongly disordered random singlet (RS) phase of the spin chain.
Through an asymptotically exact RG scheme closely related to that of 
Dasgupta and Ma,\cite{mada} he showed that the RS phase is characterized by a 
dynamic critical exponent $z=\infty$, meaning that the characteristic
time scale  is not a power of the charateristic length scale, but rather
an exponential function, giving so-called activated dynamics. The RS phase has  
no energy gap and the spin correlation function between two given sites vary over
several orders of magnitude from one disorder configuration to the next. The
average spatial decay of the spin correlation functions are, however,
described by power laws\cite{fish2}. Due to the
broad distributions, the typical correlations (the
average of the logarithm of the correlation function) 
behave very differently from the average correlation
functions. No string order exists. In the RS phase all the 
spins pair up and form singlets over arbitrarily large distances.
Average correlation functions are dominated by these strong pair-wise
coupling of spins, and due to the singlet nature of
the pairing, all components of the correlation functions
are predicted to decay with the same exponent, even if the
underlying Hamiltonian is not rotationally invariant. 
If a weak dimerization is now enforced the system is driven to a 
random dimer (RD) phase which is gapless, like the RS phase, but 
which has a non-vanishing string order\cite{hyma}, like the Haldane phase.

Recently Young and Rieger\cite{youn} were able to numerically verify 
many of these 
striking RG predictions by using a mapping to free fermions.
In this paper we have extended their results to larger system sizes
and we have studied additional correlation functions in the RS and
RD phases.  Different components of the correlation functions
are found to decay with the same exponents, in agreement with theoretical
RG predictions, but contrary to earlier numerical work\cite{rode}.
In the RD phase we present the full finite size scaling cross-over functions
for the average and typical values of both components of the spin and string
correlation functions.

\section{Model}
We consider the following Hamiltonian $H$ of the XX chain:
\begin{equation}
 H=\sum^N_{i=1}J_i(S_i^xS_{i+1}^x+S_i^yS_{i+1}^y ) 
\end{equation}
where $S_i^{\alpha}$ are spin-1/2 operators obeying periodic
boundary conditions ($\vec S_{N+1}=\vec S_1$) and  $J_i$ are positive 
coupling constants.

In the disordered system we present results for
the following flat bond distributions:
\begin{equation}
\begin{array}{ccc}
\nonumber P(J_o)&=&(2+d)^{-1}\theta(2+d-J_o)\theta(J_o),\\
 P(J_e)&=&(2-d)^{-1}\theta(2-d-J_e)\theta(J_e), 
\label{flat}
\end{array}
\end{equation}
where $J_o$ and $J_e$ denote odd and even bonds. The dimerization
is defined as
\begin{equation}
\delta={\lbrack\ln J_o\rbrack_{\hbox{av}}-\lbrack\ln J_e\rbrack_{\hbox{av}}
\over \hbox{var}\lbrack\ln J_o\rbrack + \hbox{var}\lbrack\ln J_e\rbrack},
\end{equation}
where var denotes variance. In terms of the above distributions this 
reduces to
\begin{equation}
\delta={1\over 2} \ln({2+d\over 2-d})
\label{dime}
\end{equation}
In the disorder free dimerized system $J_o=1-d/2$ and $J_e=1+d/2$ and we
use Eq. (\ref{dime}) to define the dimerization $\delta$.

The longitudinal component of the spin correlation function, $C^z(r)$, is 
defined as
\begin{equation} 
	C^z(r)={4\over N} \sum^N_{i=1}\langle S^z_iS^z_{i+r}\rangle.
\end{equation}
The transverse component, $C^x(r)$, is defined analogously.

The longitudinal component of the string correlation function, $O^z(r)$,
is defined as
\begin{eqnarray} 
\nonumber	O^z(r)= -{4\over N} \sum^N_{i=1}\langle S_i^z 
	\exp\lbrack i\pi ( S^z_{i+1} + S^z_{i+2}+\cdots\\
	+ S^z_{i+r-1})\rbrack
	S_{i+r}^z \rangle.
\end{eqnarray}
Using the identity $S^z=\exp (i\pi S^z)/2i$ this can be rewritten,
for spin-1/2 operators and r odd as
\begin{equation} 
	O^z(r)={2^{r+1}\over N} \sum^N_{i=1}\langle 
	S_i^z S_{i+1}^z \cdots S_{i+r}^z \rangle.
\end{equation}
The behavior of the string correlation function in a dimerized system
depends on whether the site index $i$ in the sum above is odd
or even. Therefore we have separated the above expression into
a sum over even sites, defining  $O_e^z(r)$ as
\begin{equation} 
	O_e^z(r)={2^{r+2}\over N} \sum_{i \hbox{ even}}
\langle S_i^z S_{i+1}^z \cdots S_{i+r}^z \rangle,
\end{equation}
and a sum over odd sites, defining $O_o^z(r)$ as
\begin{equation} 
	O_o^z(r)={2^{r+2}\over N} \sum_{i \hbox{ odd}}
\langle S_i^z S_{i+1}^z \cdots S_{i+r}^z \rangle.
\end{equation}
The difference between the the expressions is denoted
$O^z_d(r)=O^z_o(r)-O^z_e(r)$. The transverse component of the 
string order function $O^x$ is defined analogously.

This correlation function was introduced \cite{nijs,arov,kohm,hida} to 
measure hidden long-range correlations in integer spin chains where the 
ordinary spin-spin correlation function vanishes exponentially. 
If we, for example 
have a perfectly dimerized spin chain with $J_{2i}=\infty$ and 
$J_{2i+1}=0$, then the spins form singlets
around the strong bonds, and neighboring singlets are uncorrelated,
with $O^z_d=O^z_o=1$ and $O^z_e=0$. If, on the other hand there is no 
dimerization in the system $O^z_e=O^z_o=O^z$ and hence $O^z_d=0$.
The string order parameter $O_{str}^z$ is defined as
\begin{equation}
 O_{str}^z=\lim_{r\to\infty}O_o^z(r)=
	    \lim_{r\to\infty}O_d^z(r).
\end{equation}
We are, however, working with finite size systems with periodic
boundary conditions and all results are calculated at the largest
distance around the ring. Results for $C^x(r)$ are calculated at $r=N/2$.
$C^z(r)$ vanishes for r even (see Appendix) and so results are for
$r=N/2-1$. Both components of the string order function are
defined for odd r, and they are calculated at $r=N/2-1$, using
our definition of the correlation functions. For 
simplicity we have, however, plotted all results as a function
of $n=N/2$. This will introduce some corrections to scaling
for $C^z$ and will be discussed below.

All decay exponents $\theta$ are defined according to
\begin{equation}
	f(r) \propto r^{-\theta}.
\end{equation}

\section{Mapping to free fermions}
Using the Jordan-Wigner transformation we can map the XX model
onto free fermions \cite{frad}. Therefore  we only need to solve
the one-body problem which involves diagonalizing a $(N\times N)$
matrix and hence we can study fairly large system sizes, up to
about $N=2048$ for a clean system, and up to $N=1024$ when disorder
averages are taken over several thousand configurations.

We apply the Jordan-Wigner transformations
\begin{equation}
\begin{array}{ccc}
S^-_i&=&\displaystyle\exp(-i\pi\sum_{j=1}^{i-1}c^{\dagger}_jc_j)c_i\\
S^+_i&=&\displaystyle c^{\dagger}_i\exp(i\pi\sum_{j=1}^{i-1}c^{\dagger}_jc_j)
\end{array}
\end{equation}
to our Hamiltonian and obtain
\begin{equation}
H=\sum_{i=1}^N{J_i\over 2}\lbrack c^{\dagger}_ic_{i+1}+
c_ic^{\dagger}_{i+1}\rbrack .
\end{equation}
Next we must calculate the string correlation function in the fermion
language:
\begin{equation}
\begin{array}{ccc}
 \prod_{j=i}^{i+r} S_j^z 
	&=& \displaystyle\prod_{j=i}^{i+r} (c^{\dagger}_j c_j -{1\over 2}) \\
        &=&\displaystyle({-1\over 2})^{r+1}  \prod_{j=i}^{i+r}  
        (1-2c^{\dagger}_j c_j) \\
        &=&\displaystyle ({-1\over 2})^{r+1} \prod_{j=i}^{i+r}  
             (c^{\dagger}_j+ c_j)(c^{\dagger}_j- c_j) 
\end{array}
\end{equation}
\noindent
We proceed by rewriting this\cite{youn,lieb} in the form
\begin{equation}
\prod_{j=i}^{i+r} S_j^z =({-1\over 2})^{r+1}\prod_{j=i}^{i+r} 
A_jB_j .
\end{equation}
where $A_j=(c^{\dagger}_j + c_j)$ and $B_j=(c^{\dagger}_j - c_j)$.
This expression is easily evaluated using Wick's theorem.
First we can show that
\begin{eqnarray} 
\langle A_iA_j \rangle&=&\delta_{ij}\\
\langle B_iB_j \rangle&=&-\delta_{ij}\\
\langle B_iA_j \rangle&=&\langle B_jA_i \rangle=-\delta_{ij}+2
\langle c^{\dagger}_jc_i\rangle \\
\langle A_jB_i \rangle&=-&\langle B_iA_j \rangle=\delta_{ij}-2
\langle c^{\dagger}_jc_i\rangle
\end{eqnarray}
and the only non-zero contractions to appear are then of the form
$\langle A_jB_i \rangle$ or $\langle B_jA_i \rangle$.
This leads to the following determinant:
\begin{eqnarray}
\nonumber \langle \prod_{j=i}^{i+r} S_j^z \rangle=({-1\over 2})^{r+1}
\times \\ 
\left|\begin{array}{ccccc}
\langle A_iB_i\rangle & \langle A_iB_{i+1}\rangle & . 
& \langle A_iB_{i+r}\rangle \\
\langle A_{i+1}B_i\rangle & \langle A_{i+1}B_{i+1}\rangle & . 
& \langle A_{i+1}B_{i+r}\rangle \\
     .     &       .        & . &   .      \\  
\langle A_{i+r}B_i\rangle & \langle A_{i+r}B_{i+1}\rangle & . 
& \langle A_{i+r}B_{i+r}\rangle 
\end{array} \right|.
\end{eqnarray}
Care has to taken since the fermion creation and annihilation
operators obey periodic boundary conditions when $N/2$
is odd and antiperiodic boundary conditions when $N/2$
is even.

All that remains to do is to calculate $\langle A_iB_j \rangle$.
We find the unitary transformation $U$ that diagonalizes the Hamiltonian:
\begin{equation}
\begin{array}{ccc} 
d^{\dagger}_i=\sum_jc_j^{\dagger}U_{ji} &\qquad& 
c^{\dagger}_i=\sum_jd_j^{\dagger}U_{ij}\\
d_i=\sum_jc_jU_{ji} &\qquad& 
c_i=\sum_jd_jU_{ij},
\end{array}
\end{equation}
where $c_i^{\dagger}$ denotes a fermion creation operator in the old 
basis and $d_i^{\dagger}$ denotes a creation operator in the new basis.
Then we find that
\begin{equation}
 \langle c^{\dagger}_ic_j\rangle = 
   \langle \sum_k d^{\dagger}_kd_kU_{ik}U_{jk}\rangle, 
\end{equation}
and in the ground state
\begin{equation}
 \langle c^{\dagger}_ic_j\rangle = \sum_{k<k_F}U_{ik}U_{jk}. 
\end{equation}
Hence
\begin{equation}
\langle A_iB_j\rangle= \delta_{ij}-2\sum_{k<k_F}U_{ik}U_{jk}.
\end{equation}

This completes the expression for the longitudinal component of the
string correlation function as a determinant. The equivalent
expression for the other correlation functions are obtained
in a very similar way\cite{lieb} and we only state the results. 
The transverse component of the string correlation function is
\begin{eqnarray}
\nonumber \langle \prod_{j=i}^{i+r} S_j^x \rangle=({-1\over 2})^{r+1} 
\times \\
\left|\begin{array}{ccccc}
\langle B_iA_{i+1}\rangle & \langle B_iA_{i+3}\rangle & . 
& \langle B_iA_{i+r}\rangle \\
\langle B_{i+2}A_{i+1} \rangle & \langle B_{i+2}A_{i+3} \rangle & . 
& \langle B_{i+2}A_{i+r}\rangle \\
     .     &       .        & . &   .      \\  
\langle B_{i+r-1}A_{i+1} \rangle & \langle B_{i+r-1}A_{i+3} \rangle & . 
& \langle B_{i+r-1}A_{i+r} \rangle
\end{array} \right|.
\end{eqnarray}

Next we turn to the longitudinal component of the spin correlations 
function, which can be expressed as
\begin{equation}
\langle S_i^zS_{i+r}^z \rangle=({1\over 4}) 
\left|\begin{array}{ccc}
\langle A_iB_i\rangle & \langle A_iB_{i+r}\rangle  \\
\langle A_{i+r}B_i\rangle & \langle A_{i+r}B_{i+r}\rangle  \\
\end{array} \right|,
\end{equation}
the transverse component is given by
\begin{eqnarray}
\nonumber \langle S_i^xS_{i+r}^x\rangle=({1\over 4}) \times \\ 
\left|\begin{array}{ccccc}
\langle B_iA_{i+1} \rangle& \langle B_iA_{i+2} \rangle& . 
& \langle B_iA_{i+r}\rangle \\
\langle B_{i+1}A_{i+1} \rangle & \langle B_{i+1}A_{i+2} \rangle & . 
& \langle B_{i+1}A_{i+r}\rangle \\
     .     &       .        & . &   .      \\  
\langle B_{i+r-1}A_{i+1} \rangle & \langle B_{i+r-1}A_{i+2} \rangle & . 
& \langle B_{i+r-1}A_{i+r} \rangle
\end{array} \right|.
\end{eqnarray}

\section{Results}
\begin{figure}
\centering
\epsfysize=6cm
\leavevmode
\epsffile{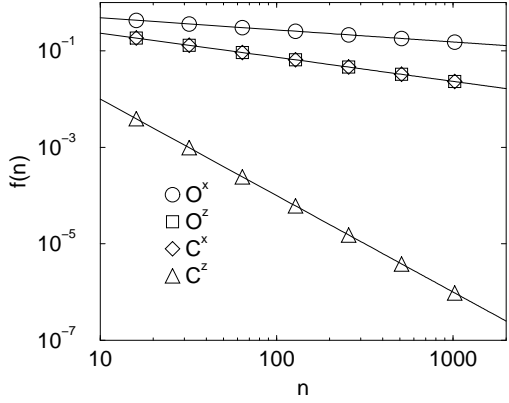}
\caption{The spin and string correlations functions
         in the clean XX model. System sizes go from $N=32$ to $N=2048$ and
	$n=N/2$. The solid lines are linear fits to the three largest systems.
	Slopes are given in Table \ref{tab01}. }
\label{fig01}
\end{figure}

In order to check the accuracy to which we can recover
the exponents of the clean system we first display all correlations 
functions for the pure model in a log-log plot in Fig. \ref{fig01}. 
In the pure model, at zero temperature,
Lieb, Schultz and Mattis\cite{lieb} showed that $C^z(r)$ is identically
zero for r even, but decays with an exponent of
2 at large r (odd). Later McCoy\cite{mcco} showed that $C^x(r)$ 
decays with an exponent equal to 1/2. For the pure
model $O^z$ will equal $C^x$ and $O^x$ is the square
root of $O^z$ (see Eq. (\ref{equal})). Hence $O^z$ and $O^x$  will
decay with exponents 1/2 and 1/4 respectively. 
By using a linear fit to the data points for
the three largest system sizes the expected decay exponents 
are recovered to five significant digits (see Table \ref{tab01}). 

\begin{figure}
\centering
\epsfysize=6cm
\leavevmode
\epsffile{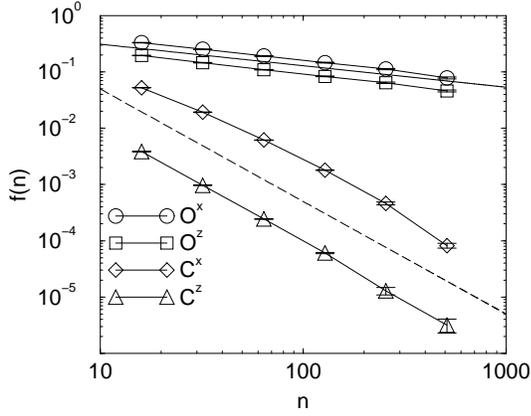}
\caption{The spin and string correlation functions
in the disordered XX model. Disorder averages are taken over 
$5\times 10^3- 10^5 $
configurations, and system sizes vary from $N=32$ to $N=1024$. The solid
and dashed lines have slopes -2 and -0.382 respectively.}
\label{fig02}
\end{figure}

We next add flat disorder, described by Eq. (\ref{flat}),
and expect to see the RS exponent\cite{fish2} 2 for the
spin correlations $C^x$ and $C^z$. $O^x$ maps on to the transverse 
field Ising model
correlator, which falls of with an exponent of 0.382 
in the RS phase\cite{fish1}. We therefore expect
both $O^x$ and $O^z$ to fall off with exponent 0.382
We plot the results in Fig. \ref{fig02}.

$C^z$ has previously been reported\cite{rode,haas} to show algebraic
decay with an exponent of 2, as predicted by the RS
calculation, and we confirm this result. While
$C^z$ remains the same in the clean and RS case, the exponent of the
transverse correlation function $C^x$ is predicted to change
from 1/2 in the clean case to 2 in the RS
phase. Previous numerical results\cite{rode} for $C^x$
indicated a transition to exponential decay.
Contrary to this, our results in Fig. \ref{fig02} do support an exponent
of 2 in the RS case. The exponent appears to be slightly smaller than 2 
for the smaller system sizes, but the slope gets very close to 2 for the
larger system sizes. We believe that we do not see
the correct exponent for smaller system sizes because
$C^x$ has not yet converged to its thermodynamic limit.
The last data point ($N=1024$) is not reliable, since for this 
large system size and strong disorder 
the numerical routines have begun to become unstable, and we do not 
trust the accuracy of this point.

We believe that the exponential decay seen previously was caused 
by the use of a gaussian distribution of bonds. As the disorder gets strong 
enough this will introduce ferromagnetic bonds into the system.
The ferromagnetic bonds will destroy the antiferromagnetic
order leading to the RS phase. Westerberg et al.\cite{west1}
have shown that the spin 1/2 AF fixed point is unstable
towards introduction of FM bonds. Furthermore they
showed\cite{west2} that the presence of FM bonds leads to exponential
decay of the spin correlation function.

We note that although the decay exponent is the same for both
components of the correlation functions, the prefactors
differ. If the spins were coupled together in pairs as true
singlets this would, of course, not be the case.  
We believe that the different prefactors are caused by 
residual fluctuations of the valence bonds.

During the course of this work we did fairly extensive studies
of various other bond distributions. In particular we worked
with a power law distribution
\begin{equation}  
P(J)=\alpha J^{-1+\alpha}\theta(J)\theta(1-J). 
\end{equation}
The motivation for
a power law distribution comes from the decimation RG
procedure, in which strong bonds in the system form 
singlets and thereby generate weak induced couplings between
their neighbors. This procedure, in effect, eliminates strong bonds
and replace them by much weaker bonds. The energy scale is thus
lowered and the procedure becomes exact in the low temperature
limit\cite{fish1,hyma}. After renormalization the bond distribution
is given by the above power law distribution. We believed that
by starting as close to the fixed point distribution as possible
it would be easier to observe the predicted RS behavior of the
correlation functions. Another distribution we studied
was a flat distribution with mean equal to one, but width
less than two. With neither of these distributions did we
observe the RS behavior as clearly as with the distribution
presented in this paper. There seems to be two competing factors
that influence the results. The power law distribution is very
close to the fixed point distribution, but the bonds get
distributed over several hundred orders of magnitude and
the numerical routines become unstable and give increasingly
unreliable results. The flat distribution with width less
than two behaves very well numerically, but it is far from 
the fixed point, and we did not manage to reach system sizes
large enough to observe the expected behavior. It appears that 
a flat distribution with mean one and width two is close
enough to the fixed point for us to see the RS exponents,
but it is not so broad that it causes problems for 
the numerical routines, except at the largest system size
(N=1024).
 
\begin{figure}
\centering
\epsfysize=6cm
\leavevmode
\epsffile{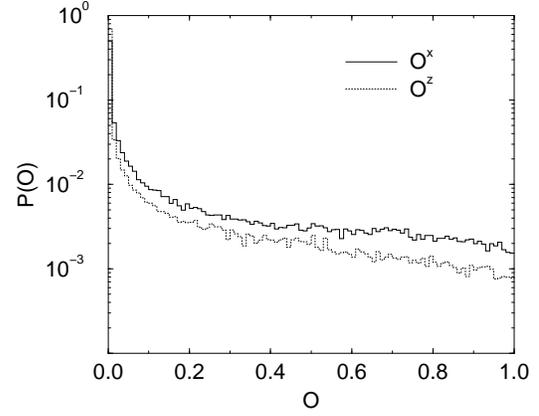}
\caption{The distribution of the transverse
and longitudinal components of the string correlation function in the RS phase
for system size $N=256$.}
\label{fig03}
\end{figure}

The string correlation functions decay as predicted in the RS
phase. The decay exponent of $O^x$ changes from 0.50
in the pure case to 0.382 in the RS phase, in agreement
with earlier observations\cite{youn}. Young and Rieger
measured the spin correlation function in the transverse-field
Ising chain, which maps onto the transverse
component of the string correlation function in the
XX model, see Eq. (\ref{corr}). The decrease in exponent
is of interest since the disorder actually causes
the correlation function to fall off slower than in the
clean case. We do, 
however, have to keep in mind that we are looking
at the average correlation function and not the
typical correlation function. The average correlation
function is dominated by a few strong couplings
between distant spins. Such strong couplings are
a characteristic of the RS phase. The typical correlation 
function, on the other hand, shows a very
different behavior\cite{fish1,fish2,fish3} and will be
studied below.
We do also see the predicted change in $O^z$, which changes from exponent
0.25 in the pure case to 0.382 in the RS phase.
$O^z$ is the square of $O^x$ and it is interesting
that the disorder average of both quantities
decay with the same exponent. Intuitively this can be understood
by considering a correlation function that 
is dominated by a few strong correlations of order 1, while
the rest of the correlations are of order 0. The actual distributions 
are plotted in Fig. \ref{fig03}.

\begin{figure}
\centering
\epsfysize=6cm
\leavevmode
\epsffile{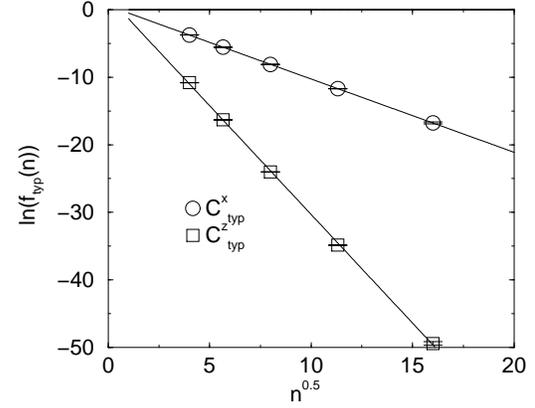}
\caption{The typical correlation
functions plotted versus the square root of the system size. The
solid lines are linear fits to all data points.}
\label{fig04}
\end{figure}

\begin{figure}
\centering
\epsfysize=6cm
\leavevmode
\epsffile{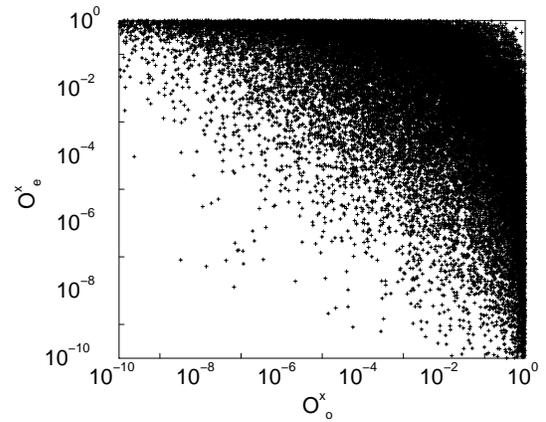}
\caption{$O_e^x$ plotted against $O_o^x$ in the RS phase for $N=256$.}
\label{fig05}
\end{figure}

The typical correlation function, here obtained by
exponentiating the disorder average of the logarithm
of the correlation function is predicted\cite{fish1} to
decay according to
\begin{equation}
f_{\hbox{typ}}(r) \propto \exp(-A\sqrt{r}),
\end{equation}
where A is some nonuniversal constant. By plotting
the log of $C_{\hbox{typ}}^x$  and 
$C_{\hbox{typ}}^z $ against $\sqrt{r}$ in Fig. \ref{fig04} we see that this
is indeed the case, and a linear fit to all data points gives the
value -1.08 and -3.22 respectively for the nonuniversal
constant A above. 
It is of interest to note that $C^x$ and $O^z$ are identical for the pure 
model, but in the RS phase $C^x$ and $O^z$
decay with different exponents. $C_{\hbox{typ}}^x$ and $O_{\hbox{typ}}^z$ are,
however, again equivalent in the RS phase. This is easily
understood by the mapping to the Ising chain. Looking at
Eq. (\ref{equal}) it is clear that $C^x$ and $O^z$ differ
in the RS phase
since $O_o^x$ and $O_e^x$ are anti-correlated. This is illustrated
in  Fig. \ref{fig05}
and provides further evidence for the RS phase since in the RS 
phase strong bonds can never
cross each other, and this means that if $O_o^x$
is of order unity, then $O_e^x$ is bound to be small.
The typical correlations $C^x_{\hbox{typ}}$ and $O^z_{\hbox{typ}}$ are
equivalent since
\begin{equation}
\ln(C^x(N/2))=\ln(O_o^x(N/2-1)) + \ln(O_e^x(N/2-1)),
\end{equation}
and because if there is no dimerization in the system
the average of $\ln O_o^x(i-j)$ will equal the average
of $\ln O_e^x(i-j)$.
We note that since $O^z=(O^x)^2$ the logarithms of the typical correlations
$O^z_{\hbox{typ}}$ and $O^x_{\hbox{typ}}$ will differ by a factor 2.

\begin{figure}
\centering
\epsfysize=6cm
\leavevmode
\epsffile{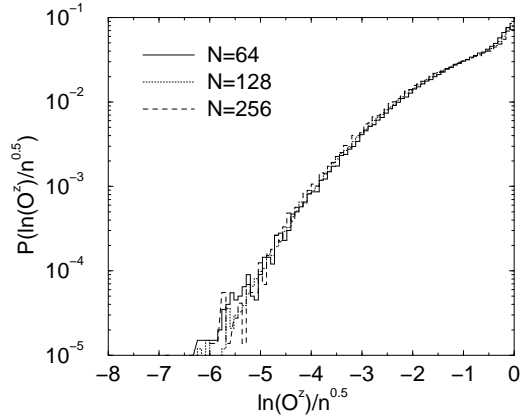}
\caption{The distribution of the longitudinal component
of the string correlation function in the RS phase.}
\label{fig06}
\end{figure}

\begin{figure}
\centering
\epsfysize=6cm
\leavevmode
\epsffile{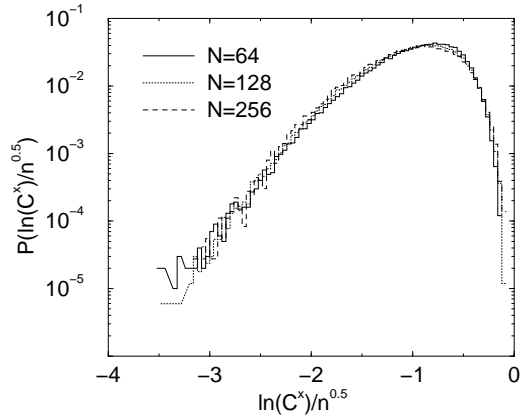}
\caption{The distribution of the transverse component
of the spin correlation function in the RS phase.}
\label{fig07}
\end{figure}

The distributions of the logarithm of the correlation functions
\begin{equation}
\ln(f(r))/\sqrt{r}
\end{equation}
is supposed to scale to a fixed distribution for large
r. Young and Rieger\cite{youn} studied this distribution for $O^x$,
and we present results for $O^z$ and  $C^x$ in Fig. \ref{fig06} and 
Fig. \ref{fig07}.
As we see the distributions scale well.
The distributions of $O^z$ and  $C^x$ are very different,
and it not obvious from  looking at the plots that the
typical correlations $O_{\hbox{typ}}^z$ and  $C_{\hbox{typ}}^x$ 
turn out to be the same (though they do). 

Next we focus on a dimerized system with no disorder.
Any observable is a function of the three length
scales in the system: the system size $N$, 
the distance around the ring $r$, and the correlation
length $\xi$. We make
the standard finite size scaling hypothesis and assume
that near the critical point ($\xi=\infty,N=\infty$)
we can express our observable in terms of a scaling 
function with dimensionless arguments formed by
ratios of our length scales. Hence a correlation
function $C$ can be expressed as 
\begin{equation}
C(N,r,\xi)\propto N^{\omega}f(N/r,N/\xi),
\end{equation}
where $\omega$ is a scaling exponent and $f(N/r,N/\xi)$ 
is the scaling function. If we measure $C$ for different
system sizes, but at a fixed ratio $N/r$ and plot
$CN^{-\omega}$ vs. $N/\xi$ the curves should
collapse onto a single scaling function $f(N/\xi)$,
independently of the values of $N$ and $\xi$.
If one assumes that $N$ is large enough so that $C$
is not a function of $N$, then one can alternatively
use data for one system size $N$, and different $r$
and plot $Cr^{-\omega}$ versus $r/\xi$. Again
the curves should collapse onto a single scaling 
function $f(r/\xi)$, independently of the values of $r$ 
and $\xi$,
but only as long as the value of $C$ has reached
its thermodynamic limit. This was done by Young and Rieger\cite{youn}, 
but in this paper we present data for different system sizes at a fixed
distance half way around the ring, $r=N/2$, so that the first
argument of the scaling function $f$  is fixed at $N/r =2$.

\begin{figure}
\centering
\epsfysize=6cm
\leavevmode
\epsffile{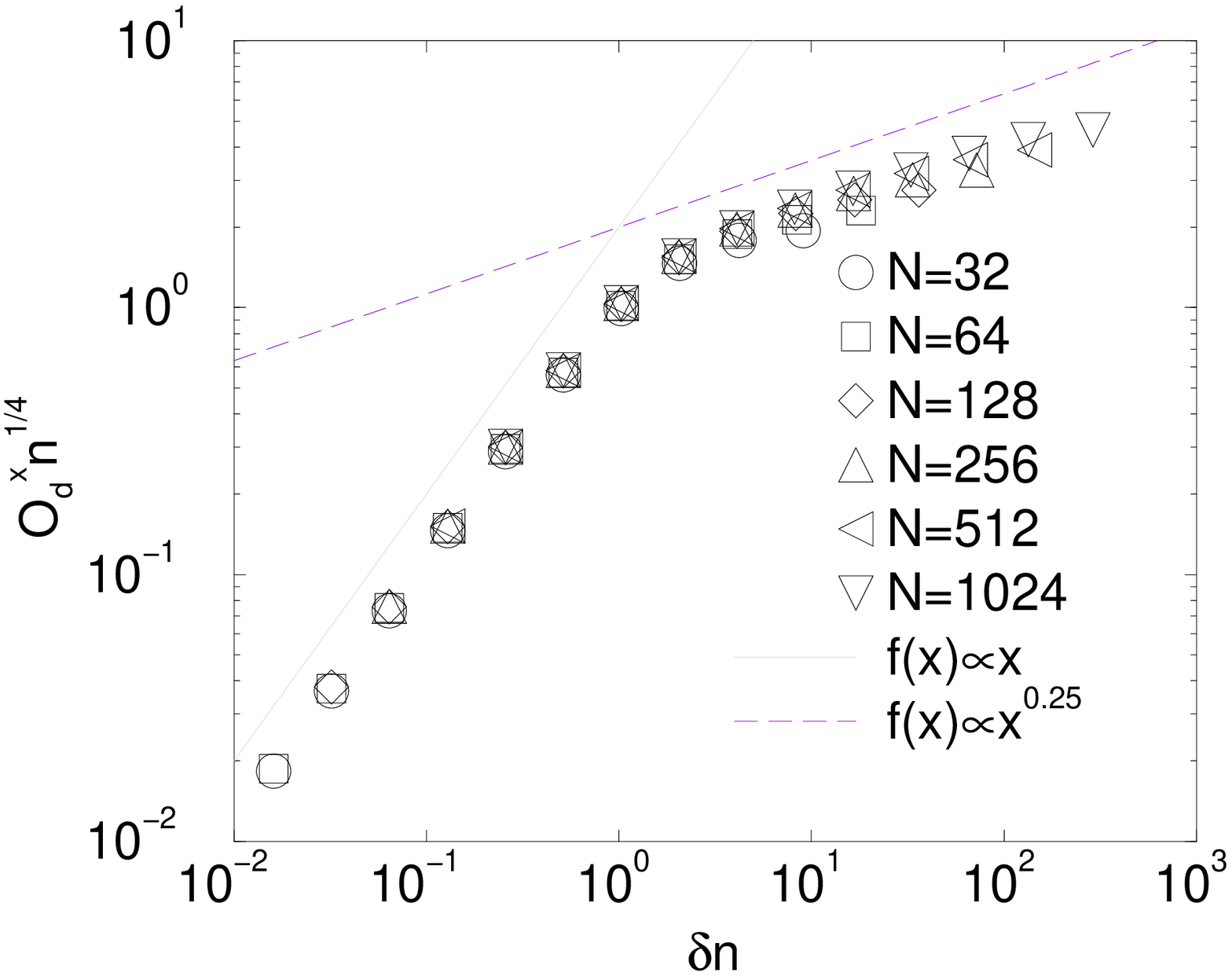}
\caption{The transverse component of the
string correlation function in the dimerized XX model.}
\label{fig08}
\end{figure}

Consider the string correlation function $O_d^x(n,\delta)$ to be a
function of the dimerization $\delta$ and $n=N/2$. 
From above we know that
\begin{equation}
O_d^x(n,\delta=0)\propto n^{-{1\over 4}}
\label{sca1}
\end{equation}
and Pfeuty\cite{pfeu} showed that (using the mapping from
the XY model to the Ising chain studied by Pfeuty)
\begin{equation}
O_d^x(n=\infty,\delta)\propto \delta^{1\over 4}.$$
\label{sca2}
\end{equation}
The correlation length
for free fermions is of the form $\xi \propto \delta^{-1}$.
Consider a string correlation function of the form
\begin{equation}
O_d^x(n,\delta)\propto n^\omega f(n\delta).$$
\end{equation}
Eq. (\ref{sca2}) requires that for asymptotically large values
of $x$, $f(x)$ behaves as $ x^{1\over 4}$,
which fixes $\omega$ to $-{1\over 4}$.
We thus expect a single
curve when $O_d^xn^{1\over 4}$ is plotted versus $n\delta$.
In Fig. \ref{fig08} a log-log plot of the scaling function shows that
this is indeed the case. For small values of
the argument the function is linear, $f(x)\propto x$, but for larger
values of the argument there is a sharp transition to the
asymptotic behavior, $f(x)\propto x^{1\over 4}$. As the
dimerization gets too big we notice that there are corrections
to scaling.

\begin{figure}
\centering
\epsfysize=6cm
\leavevmode
\epsffile{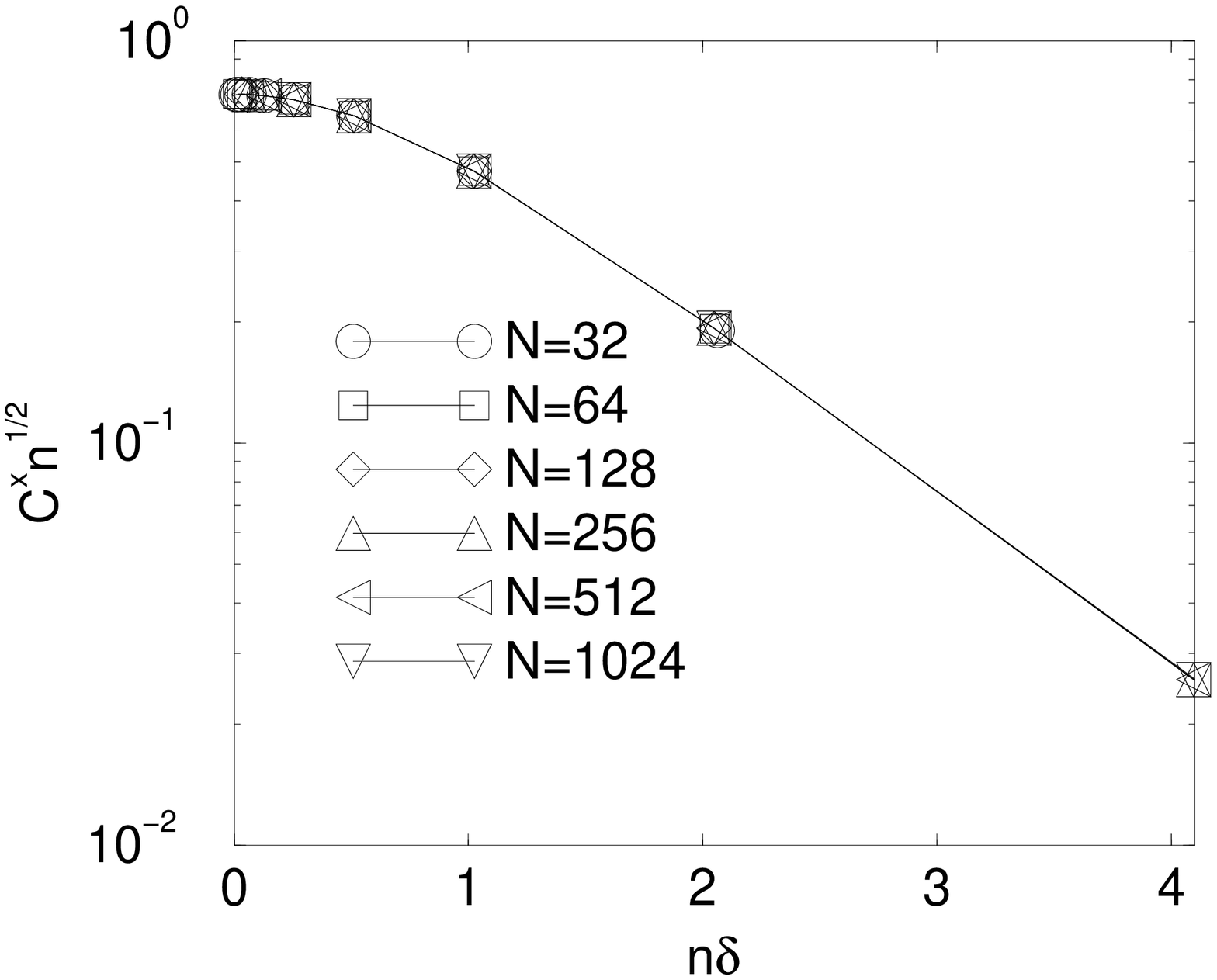}
\caption{The transverse component of the
spin correlation function in the dimerized XX model.}
\label{fig09}
\end{figure}

\begin{figure}
\centering
\epsfysize=6cm
\leavevmode
\epsffile{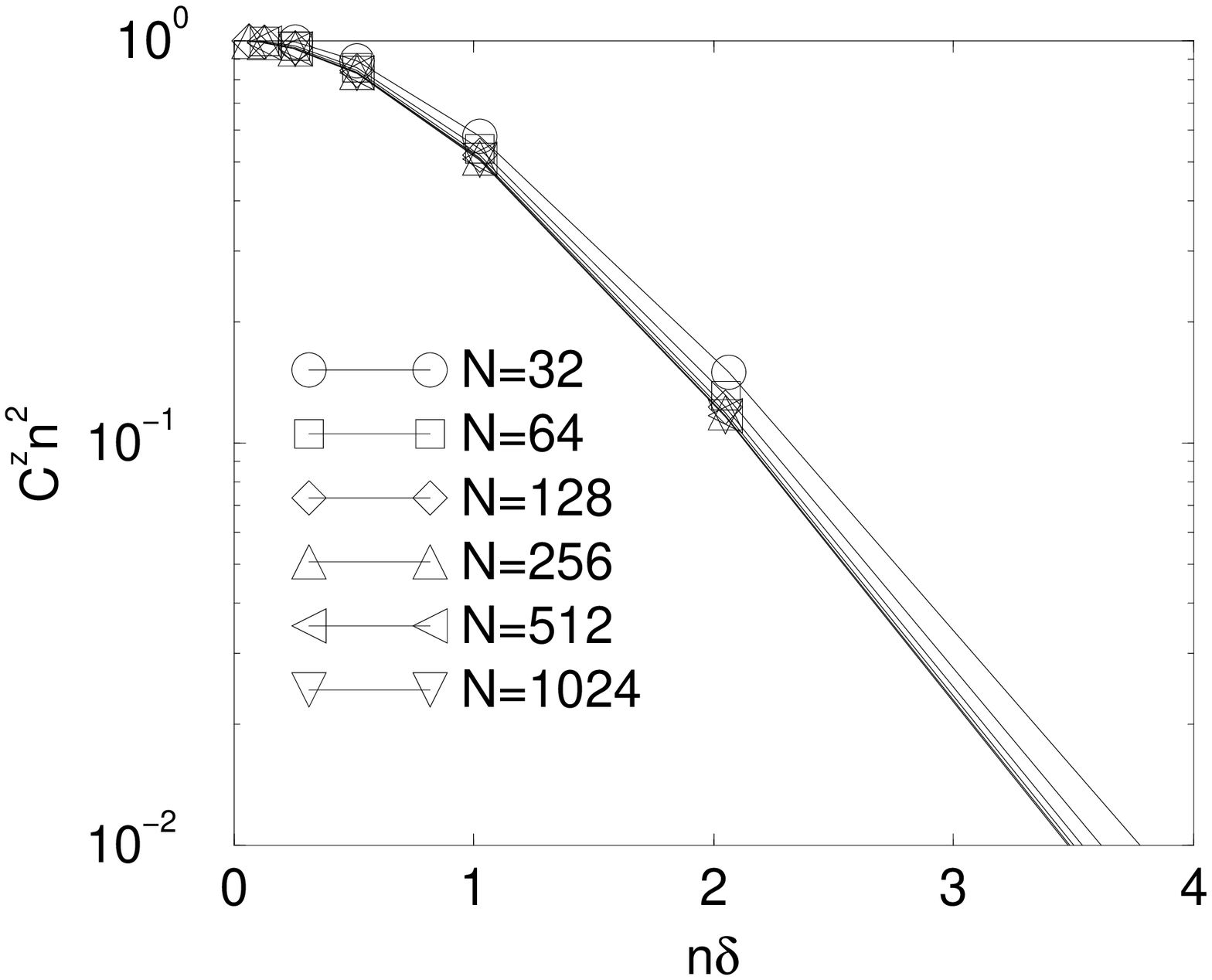}
\caption{The longitudinal component of the
spin correlation function in the dimerized XX model.}
\label{fig10}
\end{figure}

Since the longitudinal component of the string correlation
is the square of the transverse component, the above
arguments are identical for $O_d^z(N,\delta)$, except that
all exponents equal to 1/4 are changed to 1/2.
The spin correlation function $C^x$ also
scales very well with exponent 1/2, as shown in Fig. \ref{fig09}. The longitudinal
spin correlation $C^z$ shows fairly large corrections
to the scaling (see Fig. \ref{fig10}), but the corrections are
due to the fact that the correlation is measured
at $r=N/2-1$ instead of at $r=N/2$. This was checked by
by plotting the same results for system sizes with
$N/2$ odd, in which case $C^z$ can be measured at
$r=N/2$, and there are no visible correction to scaling.
But in order to be consistent throughout this paper
we have only shown data for system sizes with $N/2$ even.

If we now add disorder to the dimerized system we expect
the correlation length to be of the form\cite{fish2} 
$\xi \propto \delta^{-2}$.
Furthermore we expect that
\begin{equation}
O_d^z(n,\delta=0)\propto n^{-{0.382}}
\end{equation}
Therefore the scaling function for both components
of the string correlation function should read
\begin{equation}
O_d^{x,z}(n,\delta)\propto n^{-0.382} f(N\delta^2)$$
\end{equation}

\begin{figure}
\centering
\epsfysize=6cm
\leavevmode
\epsffile{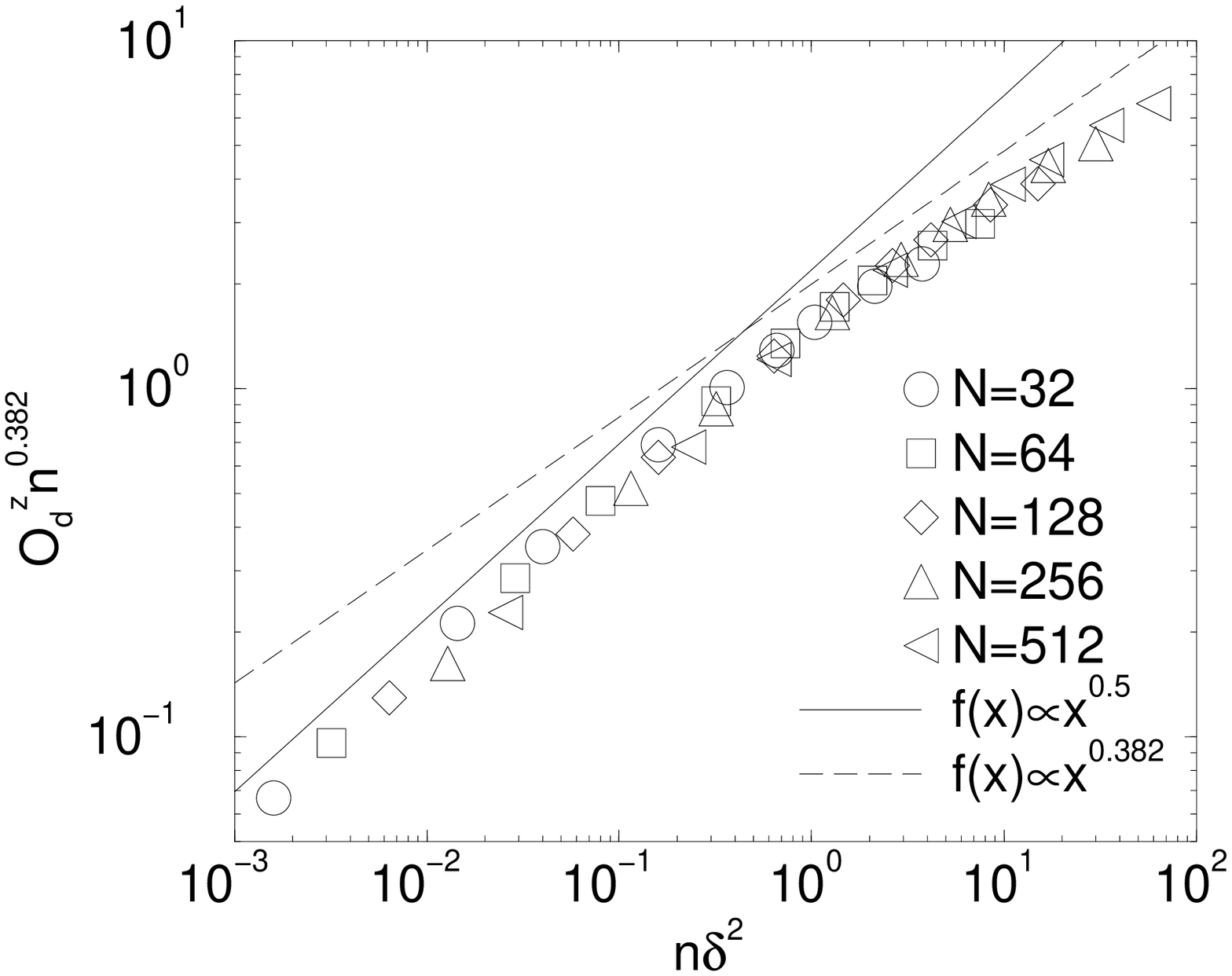}
\caption{The longitudinal component of the
string correlation function in the dimerized and disordered XX model.}
\label{fig11}
\end{figure}

\begin{figure}
\centering
\epsfysize=6cm
\leavevmode
\epsffile{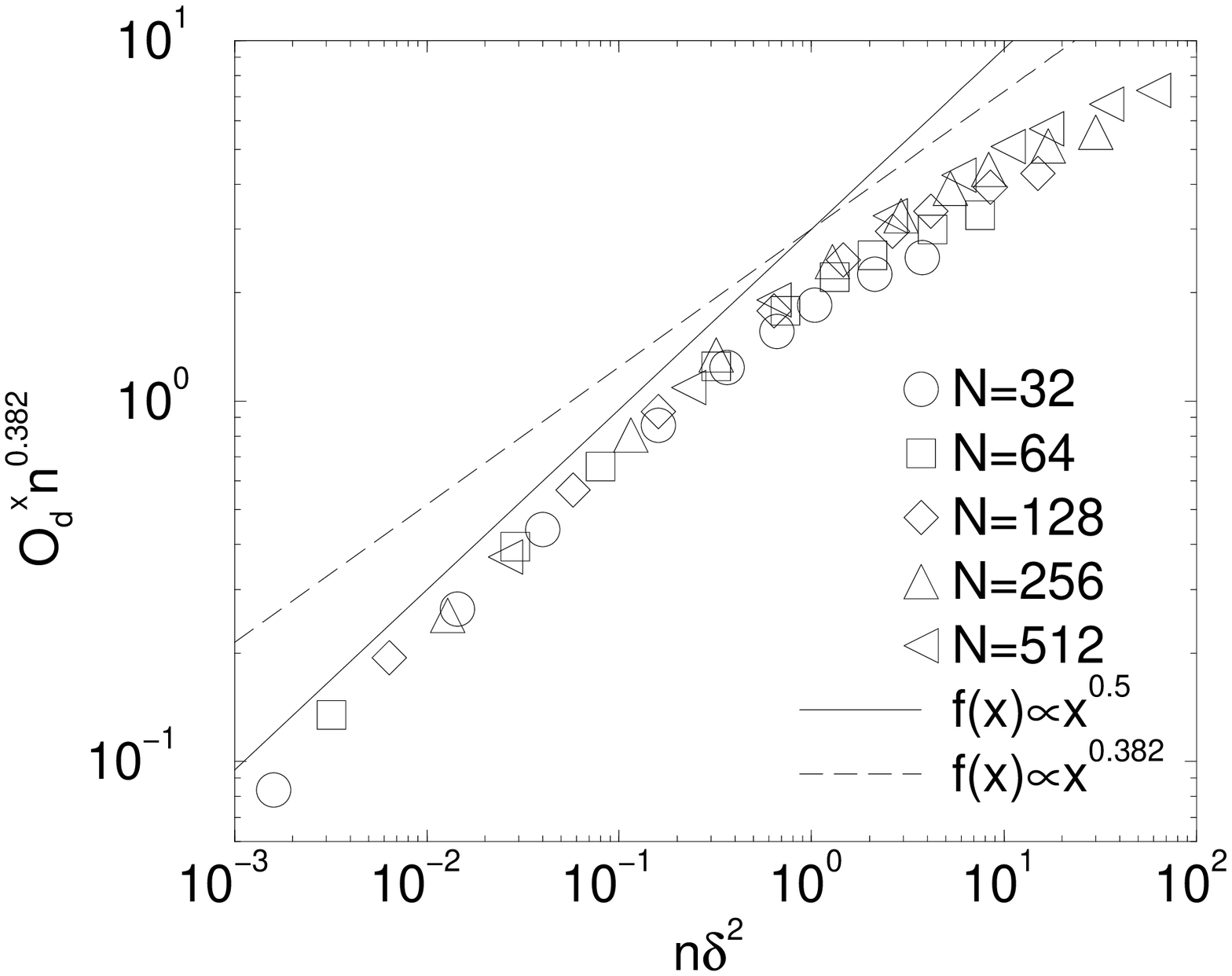}
\caption{The transverse component of the
string correlation function in the dimerized and disordered XX model.}
\label{fig12}
\end{figure}

In Fig. \ref{fig11} and Fig. \ref{fig12} we plot the scaling functions 
for $O_d^z$ and $O_d^x$. In both cases we see a transitions from
$f(x)\propto x^{1\over 2}$ to $f(x)\propto x^{0.382}$. The transition
is not as sharp as in the clean case, and it would be of
interest to have reliable data for even larger system sizes.

\begin{figure}
\centering
\epsfysize=6cm
\leavevmode
\epsffile{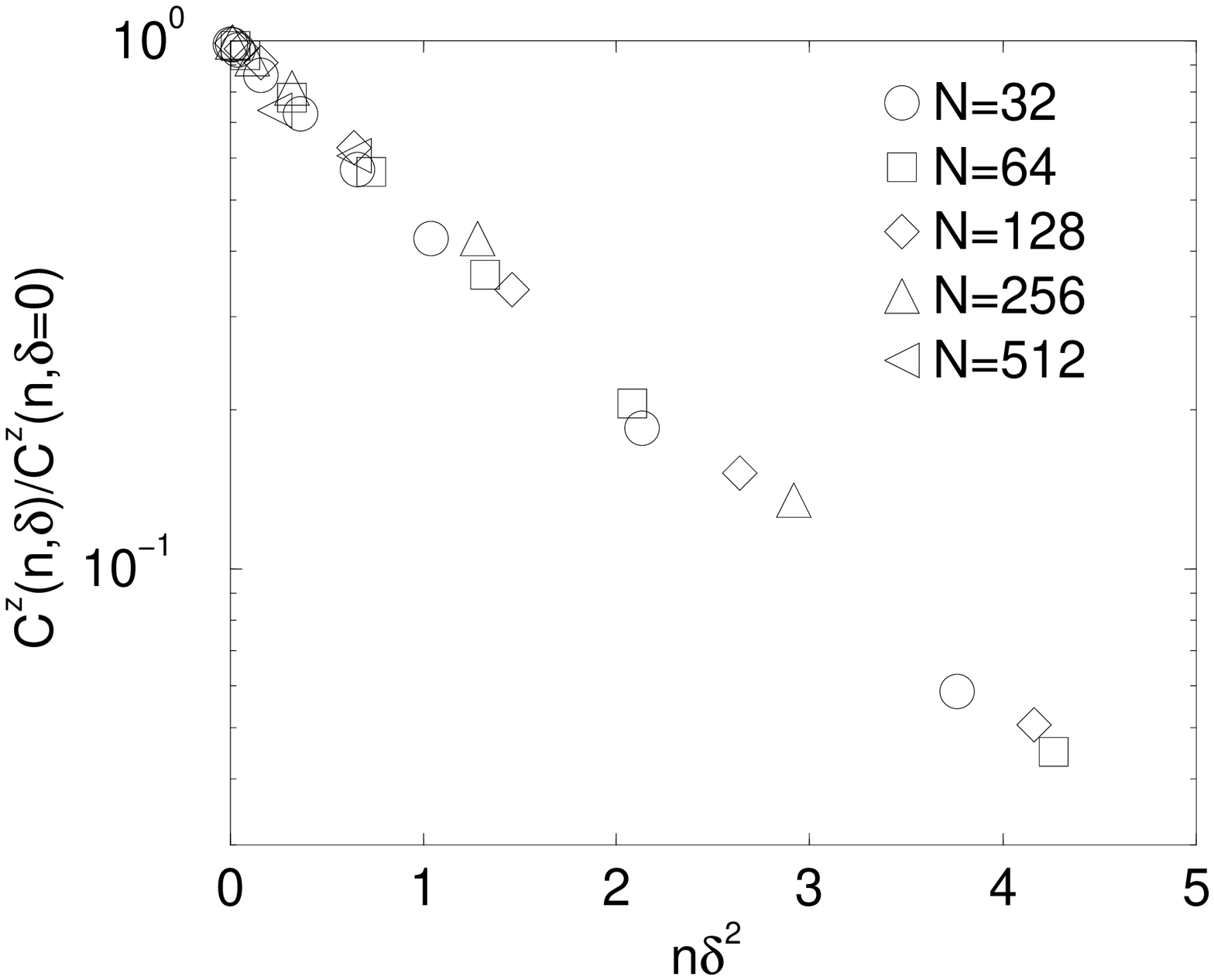}
\caption{The longitudinal component of the
spin correlation function in the dimerized and disordered XX model.}
\label{fig13}
\end{figure}

\begin{figure}
\centering
\epsfysize=6cm
\leavevmode
\epsffile{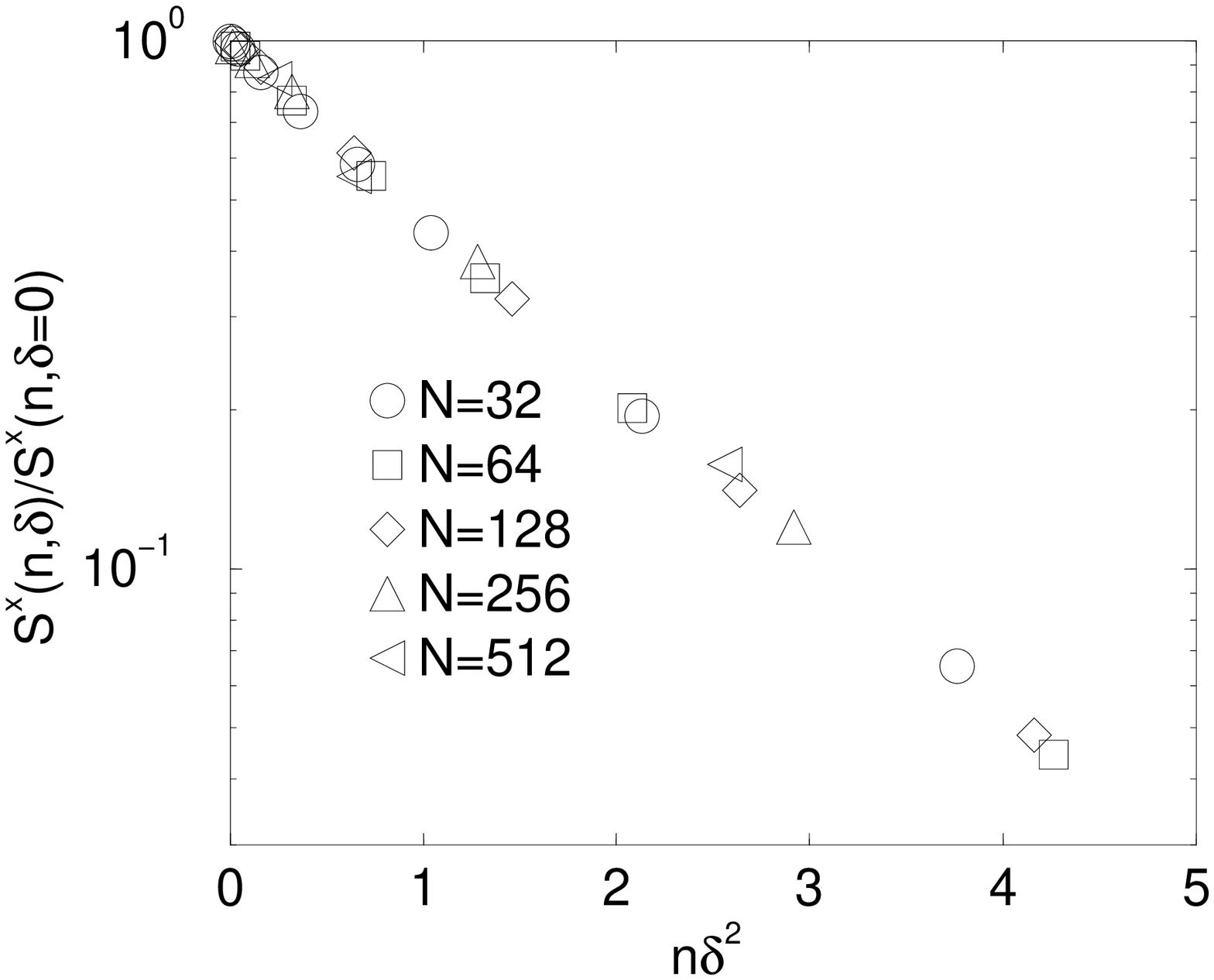}
\caption{The transverse component of the
spin correlation function in the dimerized and disordered XX model.}
\label{fig14}
\end{figure}

The average spin correlation functions should scale
according to 
\begin{equation}
C^{x,z}(n,\delta)\propto n^{-2} f(n\delta^2),$$
\end{equation}
but since the critical correlations functions have not
reached their asymptotic behavior at the system
sizes considered we compensate by plotting
$C^{x,z}(n,\delta)/C^{x,z}(n,\delta=0)$ vs. $ f(n\delta^2)$, which works
well, as can be seen in Fig. \ref{fig13} and Fig. \ref{fig14}.

\begin{figure}
\centering
\epsfysize=6cm
\leavevmode
\epsffile{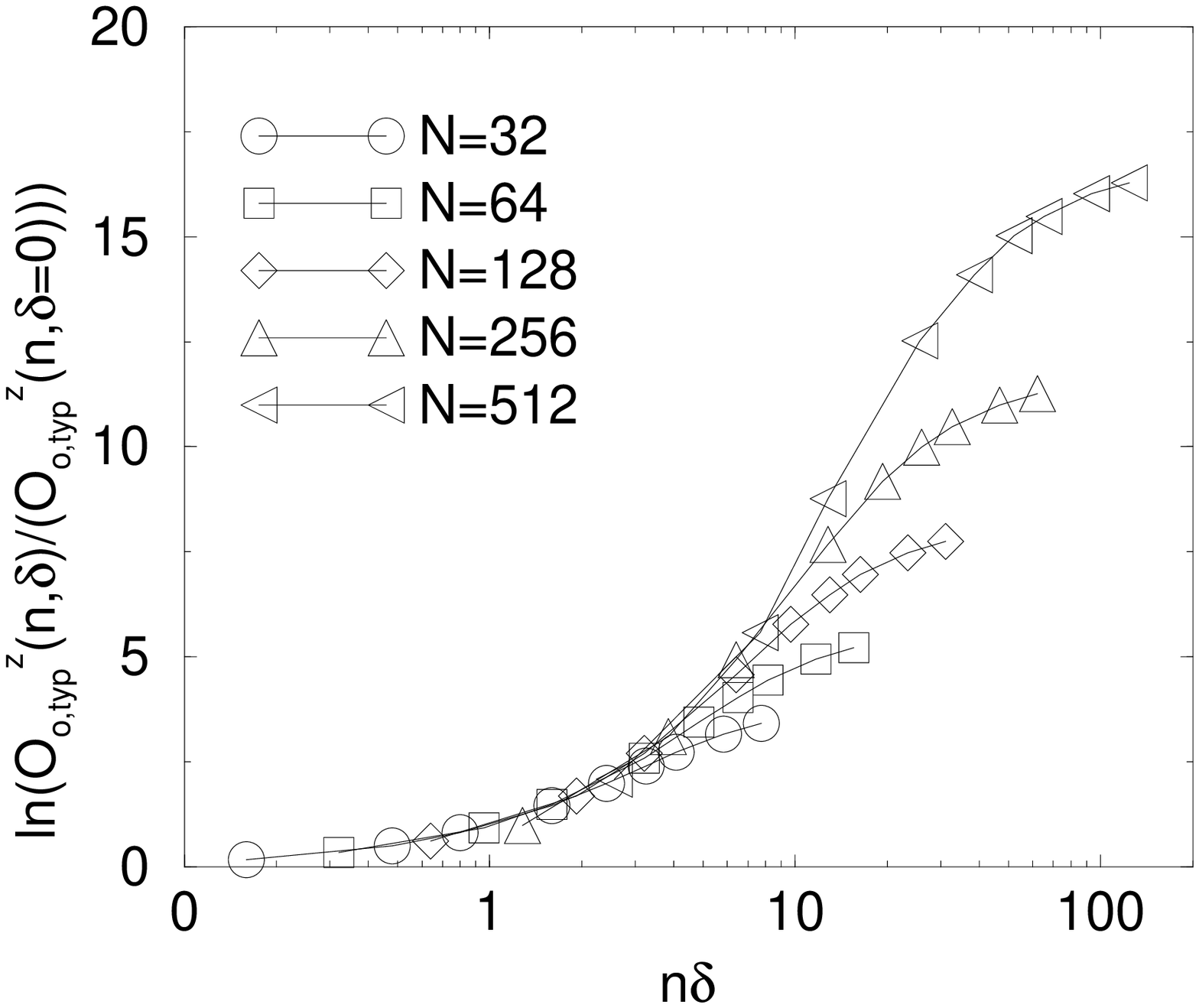}
\caption{The longitudinal component of the typical 
string correlation function in the dimerized and disordered XX model.}
\label{fig15}
\end{figure}

\begin{figure}
\centering
\epsfysize=6cm
\leavevmode
\epsffile{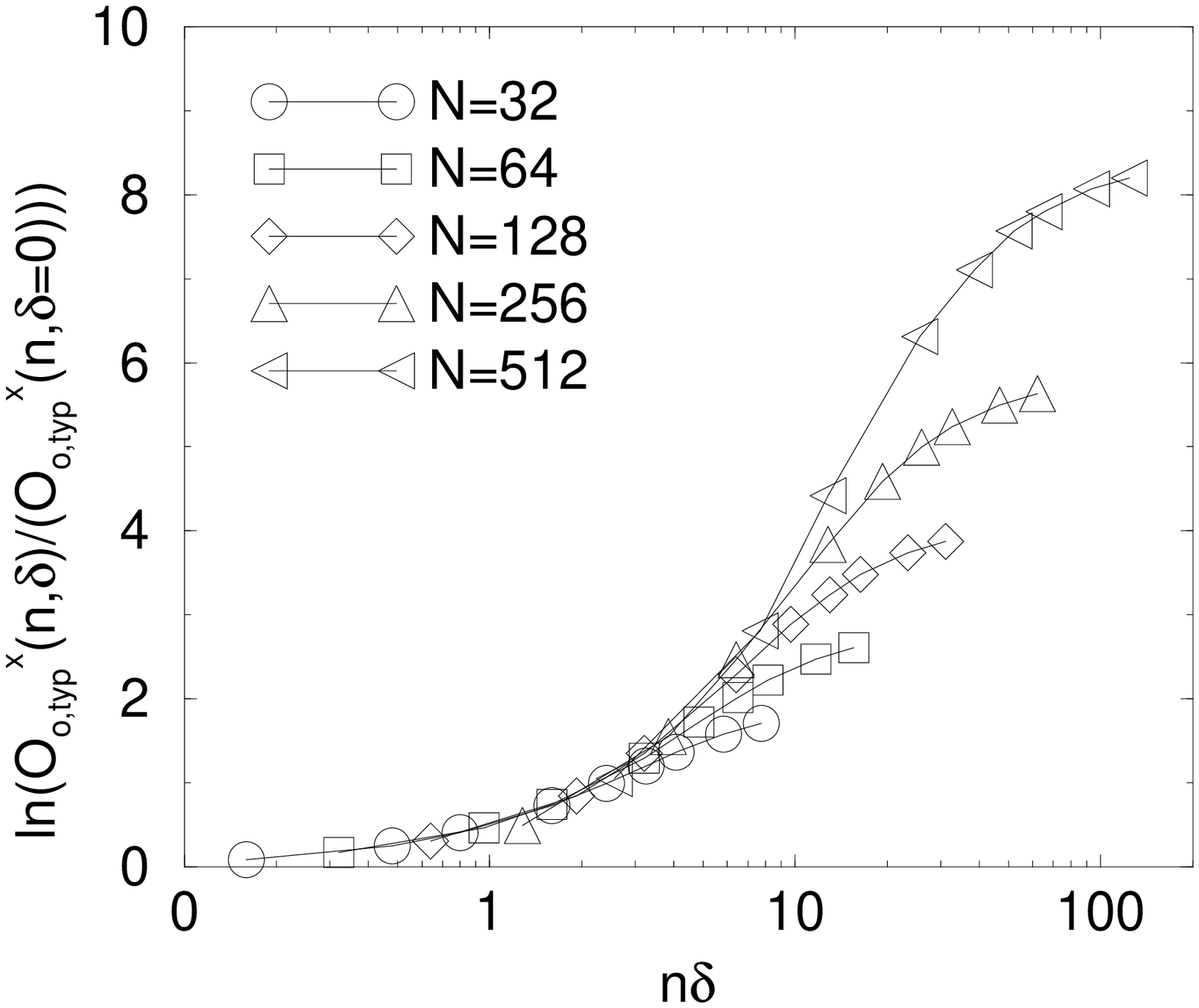}
\caption{The transverse component of the typical
string correlation function in the dimerized and disordered XX model.}
\label{fig16}
\end{figure}

\begin{figure}
\centering
\epsfysize=6cm
\leavevmode
\epsffile{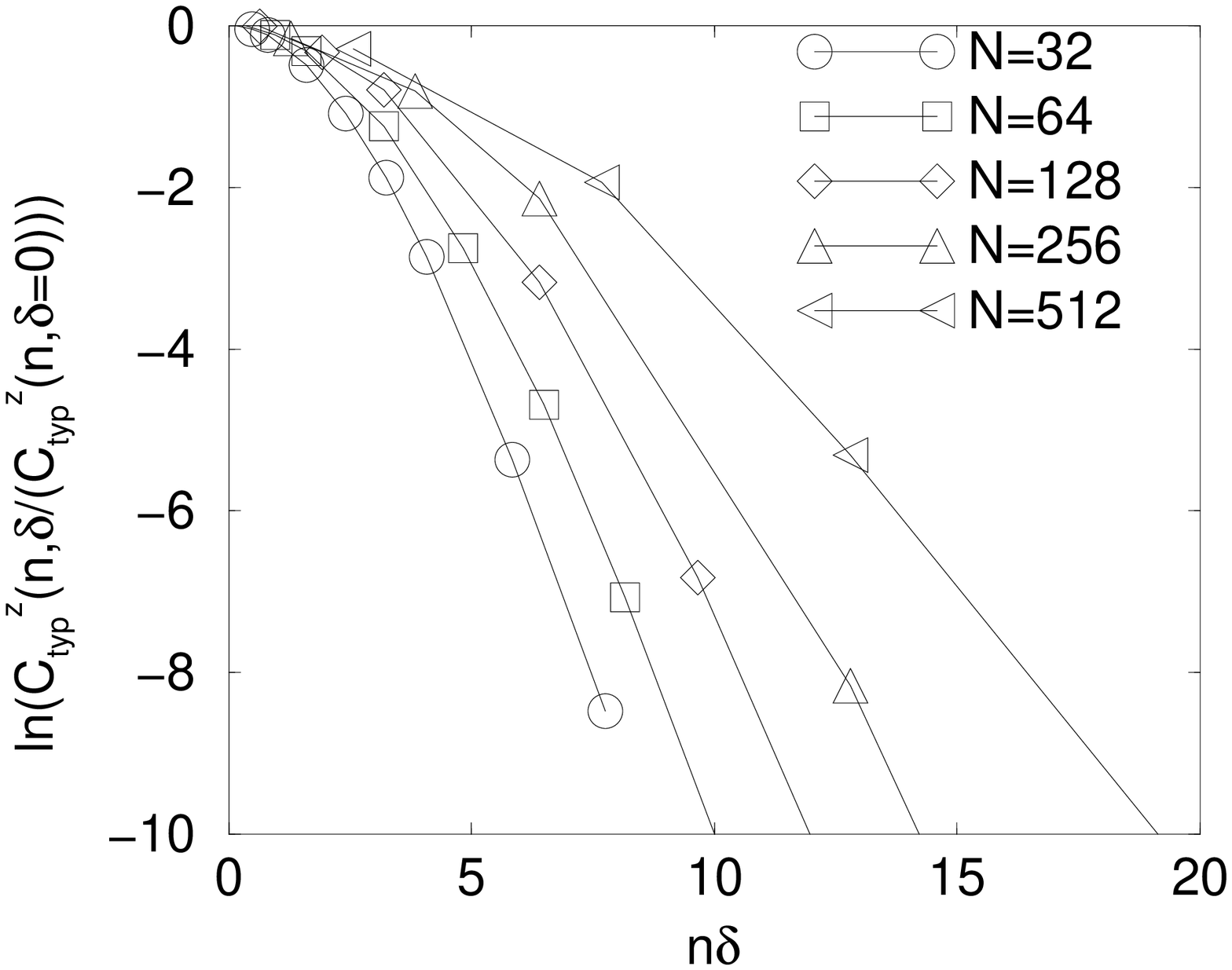}
\caption{The longitudinal component of the typical
spin correlation function in the dimerized and disordered XX model.}
\label{fig17}
\end{figure}

\begin{figure}
\centering
\epsfysize=6cm
\leavevmode
\epsffile{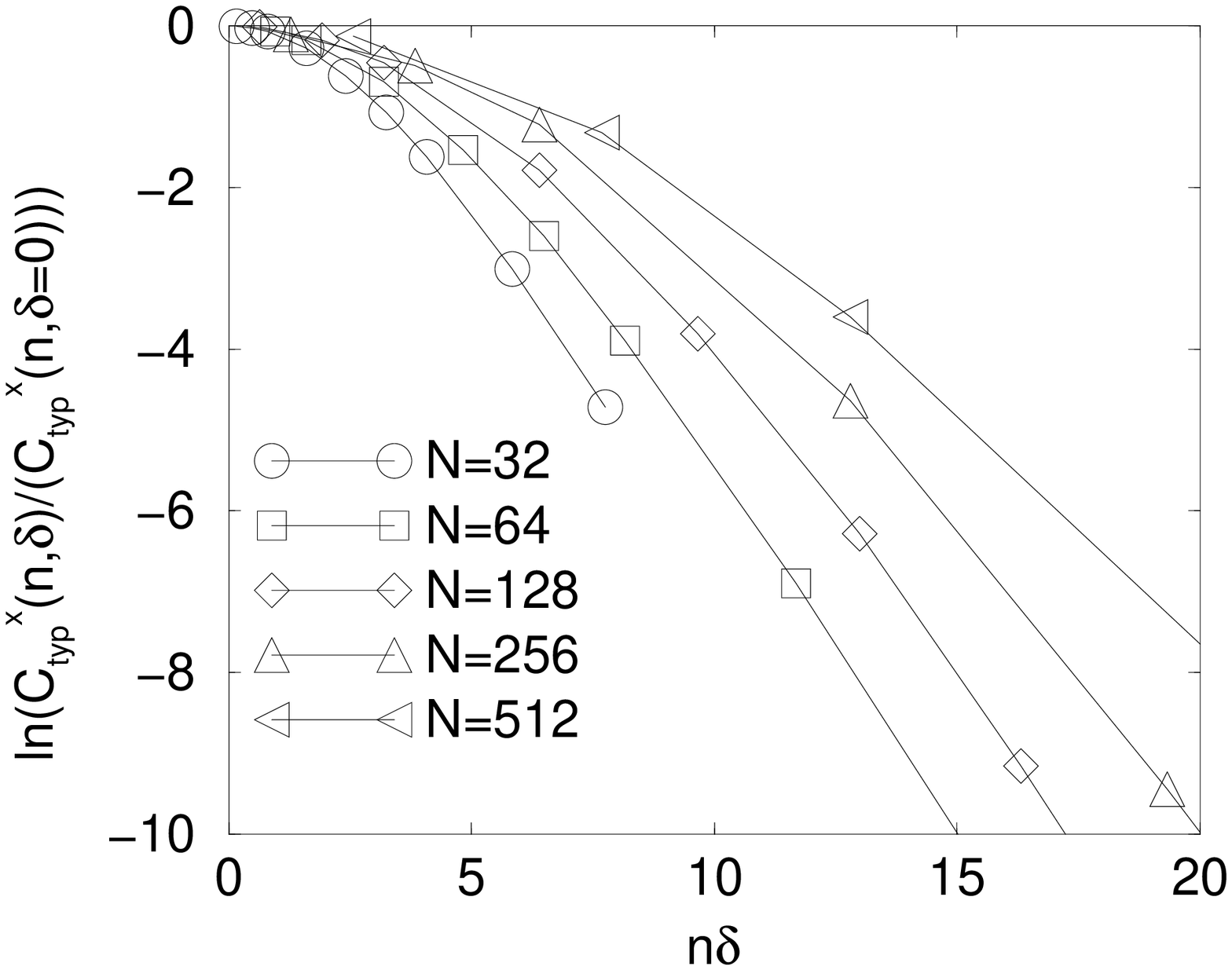}
\caption{The transverse component of the typical
spin correlation function in the dimerized and disordered XX model.}
\label{fig18}
\end{figure}

\begin{figure}
\centering
\epsfysize=6cm
\leavevmode
\epsffile{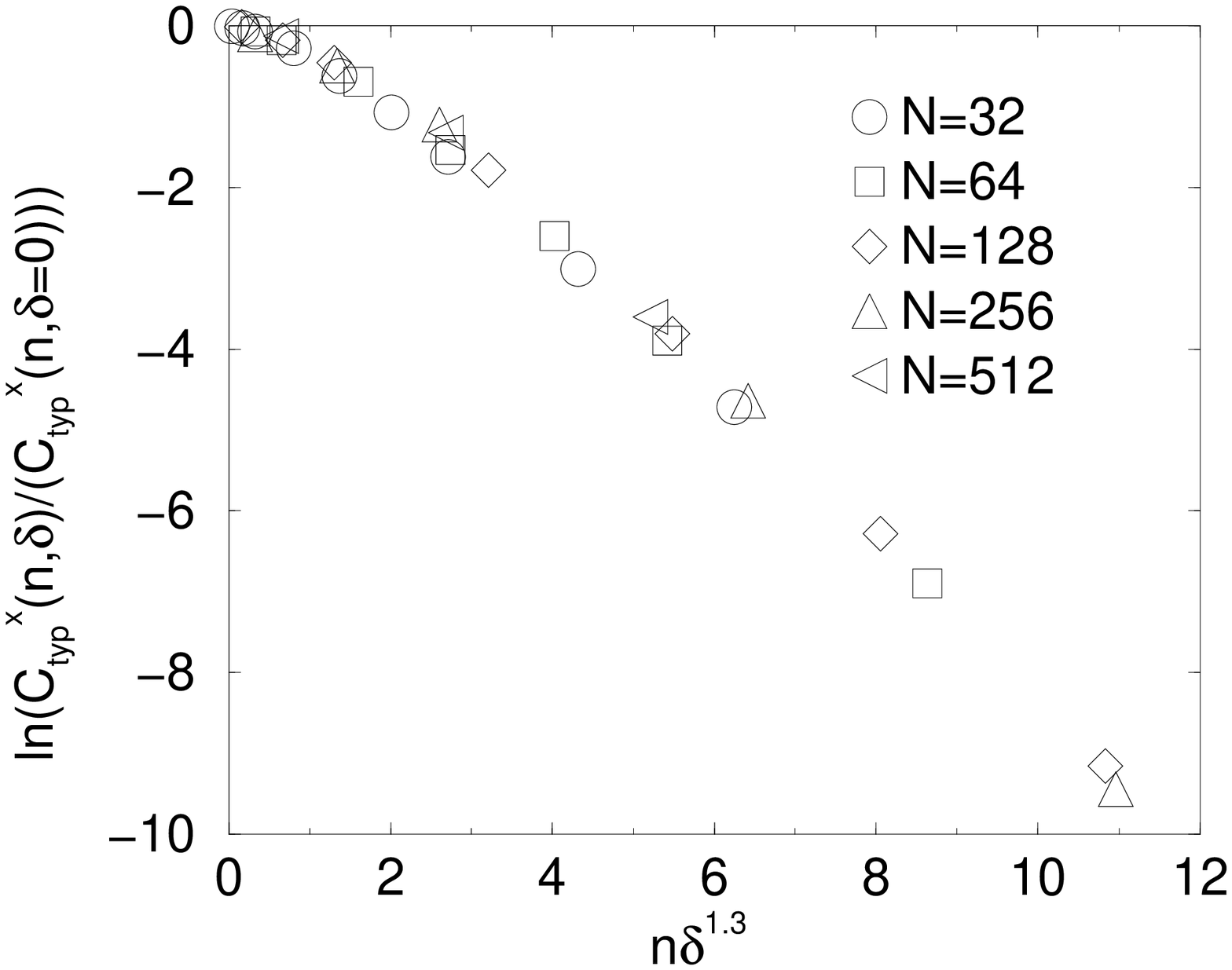}
\caption{The transverse component of the typical
spin correlation function in the dimerized and disordered XX model,
with $\xi\propto\delta^{1.3}$.}
\label{fig19}
\end{figure}

The typical correlation functions are expected to scale
as
\begin{equation}
f_{\hbox{typ}}(n,\delta)/f_{\hbox{typ}}(n,\delta=0)\propto 
f(n/\xi_{\hbox{typ}}),
\end{equation}
where $\xi_{\hbox{typ}} \propto \delta^{-1}$ in the RD phase. The string correlation
functions $O_{o,\hbox{typ}}^z$ and $O_{e,\hbox{typ}}^x$ 
appear to follow this scaling behavior fairly well, see  Fig. \ref{fig15} and Fig. \ref{fig16},
but the typical spin correlation functions in  Fig. \ref{fig17} and Fig. \ref{fig18}
show quite dramatic corrections. If we assume that 
$\xi_{\hbox{typ}} \propto \delta^{-1.3}$ the scaling works well, as shown in
Fig. \ref{fig19}. We are not sure if this discrepancy with
theory is only an effect of large corrections to scaling,
or if there is some other reason.

\section{Conclusion}
We have done extensive numerical simulations of the disordered 
XX model, using a mapping to free fermions. The decay exponents 
of the string and spin correlation functions in the RS and RD phases, 
predicted by RG calculation, have been found in very good agreement 
with numerical data. In particular the transverse component of the 
spin correlation
function is observed to decay algebraically with the correct
decay exponent, as opposed to exponentially as previously
observed\cite{rode}. We explain this difference in terms of the
bond distribution and the introduction of ferromagnetic
bonds in the previous study. We also discuss the results
of various other bond distributions, emphasizing the
importance of being close to the fixed point distribution,
but making sure that the numerical routines still
give reliable results. Full finite size scaling cross over functions
were presented for the longitudinal and transverse
components of the average and typical spin and string correlation function 
in the RD phase, as well as for the  pure dimerized
model. Only the transverse component of the
string correlation function has previously been studied
in this manner\cite{youn}. In appendix A various relations
between the different correlation functions are
derived using a mapping to the Ising model
in a transverse field. Due to the well-known
mapping to free fermions, and the use of parallel
computers, disorder averages could be taken over
system sizes up to 1024 sites, about ten times
larger than in previous studies we are aware of. 

We are extremely grateful to Ross Hyman, Senthil Todadri
and Kun Yang for their many comments and generous help. 
We acknowledge support from the NSF grant DMR 97-14055,
NSF CDA-9601632 and Ella och Georg Ehrnrooths stiftelse.

\begin{appendix}
\section{Transformation to decoupled Ising chains}
We can gain considerable insight into the behavior
of, and relationship between various correlation
functions by transforming the XY chain into
two decoupled Ising chains. Following Fisher
we use the following transformation:
\begin{eqnarray}
\nonumber \sigma_n^x&=&\prod_{j\leq n}S_j^x\\
 \sigma_n^y&=&S_n^yS_{n+1}^y\\
\nonumber\sigma_n^z&=&{2\over i}\sigma_n^x\sigma_n^y=
  {2\over i}\prod_{j\leq n}S_j^xS_n^yS_{n+1}^y.
\end{eqnarray}
The inverse transformation is given by
\begin{eqnarray}
\nonumber S_n^x&=&\sigma_{n-1}^x\sigma_n^x\\
 S_n^y&=&\prod_{j\leq n-1}\sigma_{j}^y\\
\nonumber S_n^z&=&{2\over i}S_n^xS_n^y={2\over i}\prod_{j\leq n-1}
         \sigma_{j}^y\sigma_{n-1}^x\sigma_n^x.
\end{eqnarray}
The Hamiltonian is transformed into two decoupled
Ising chains:
\begin{equation}
 H=\sum^N_{i=1}J_i(S_i^xS_{i+1}^x+S_i^yS_{i+1}^y )
  = \sum^N_{i=1}J_i(\sigma_{i-1}^x\sigma_{i+1}^x + \sigma_i^y).
\end{equation}
The two chains are dual to each other and in the pure dimerized
case one chain has coupling constant $J_o$ and transverse field
$J_e$, while the other chain has coupling constant $J_e$
and transverse field $J_o$.

Next we look at how the string correlation functions transforms:
\begin{equation}
O^x(r)=\langle S_i^xS_{i+1}^x\cdots S_{i+r}^x\rangle 
=\langle \sigma_{i-1}^x\sigma_{i+r}^x\rangle .
\end{equation}
If $r$ is even then $\sigma_{i-1}^x$ and $\sigma_{i+r}^x$ will belong
to different chains and the expectation value $<\sigma_{i-1}^x\sigma_{i+r}^x>=
<\sigma_{i-1}^x><\sigma_{i+r}^x>$ will vanish. If $r$ is odd $O^x(r)$
transforms to the Ising correlator: $O^x(r)=\langle\sigma_{i-1}^x
\sigma_{i+r}^x\rangle$.
The y component becomes (for $r$ odd):
\begin{equation}
O^y(r)=\langle S_i^yS_{i+1}^y\cdots S_{i+r}^y\rangle=
       \langle \sigma_{i}^y\sigma_{i+2}^y
       \cdots\sigma_{i+r-1}^y\rangle .
\end{equation}
The expectation values of $O^x$ and $O^y$ have to be identical
and hence
\begin{equation}
<\sigma_{i-1}^x\sigma_{i+r}^x>=
<\sigma_{i}^x\sigma_{i+2}^x\cdots\sigma_{i+r-1}^x>.
\label{eqao}
\end{equation}
This relates the expectation value of a correlation function on 
one chain to the expectation value of another correlation function
on the other chain. 

The z component of the string correlation function transforms ($r$ odd) as
\begin{equation}
O^z(r)=\langle S_i^zS_{i+1}^z\cdots S_{i+r}^z\rangle 
      =\langle \sigma_{i-1}^x\sigma_{i}^y \sigma_{i+2}^y\cdots
      \sigma_{i+r-1}^y\sigma_{i+r}^x\rangle.
\end{equation}
Using the above equality we find that
\begin{eqnarray}
\nonumber O^z(r)
&=&\langle\sigma_{i-1}^x\sigma_{i+r}^x\rangle
\langle \sigma_{i}^y\sigma_{i+2}^y\cdots\sigma_{i+r-1}^y\rangle\\
&=&\langle \sigma_{i-1}^x\sigma_{i+r}^x\rangle^2=O^x(r)^2,
\end{eqnarray}
and interestingly enough the transverse component of the string order
is simply the square of the longitudinal component.

Turning to the spin correlation function we find that
\begin{equation}
C^x(r)=\langle S^x_iS^x_{i+r}\rangle =
       \langle \sigma_{i-1}\sigma_{i}\sigma_{i+r-1}\sigma_{i+r}\rangle .
\label{corr}
\end{equation}
Hence
\begin{equation}
C^x(r)=\cases{<\sigma_{i-1}\sigma_{i+r}><\sigma_{i}\sigma_{i+r-1}>, 
		&if $r$ odd\cr
             <\sigma_{i-1}\sigma_{i+r-1}><\sigma_{i}\sigma_{i+r}>, 
                &if $r$ even.\cr}
\end{equation}
The criterion that $C^x=C^y$ gives us back Eq. (\ref{eqao}). The z component
of the spin correlation transforms to
\begin{eqnarray}
\nonumber C^z(r)=\\
\cases{<\sigma_{i-1}^x\sigma_{i+2}^y\sigma_{i+4}^y\cdots\sigma_{i+r-2}^y
        \sigma_{i+r}^x>\times \cr
             <\sigma_{i}^x\sigma_{i}^y\sigma_{i+2}^y\cdots\sigma_{i+r-2}^y
              \sigma_{i+r}^x>, 
	     &if $r$ odd\cr
<\sigma_{i-1}^x\sigma_{i+2}^y\sigma_{i+4}^y\cdots\sigma_{i+r-2}^y
               \sigma_{i+r}^x>\times \cr
             <\sigma_{i}^x\sigma_{i}^y\sigma_{i+2}^y\cdots
               \sigma_{i+r-2}^y\sigma_{i+r}^x>, 
	     &if $r$ even\cr}.
\end{eqnarray}

To summarize we want to emphasize a few important relations
between various correlation functions. These equalities follow
directly from the fact that the XY chain separates into two
decoupled Ising chains. The expressions are obtained by using the above 
transformations and assuming that $N/2$ is even, which is the case in our
simulations. The relations are:

\begin{eqnarray}
\nonumber C^x(N/2)&=&O_o^x(N/2-1)O_e^x(N/2-1)\\
O_o^z(N/2-1)&=&(O_o^x(N/2-1))^2\label{equal}\\
\nonumber O_e^z(N/2-1)&=&(O_e^x(N/2-1))^2.
\end{eqnarray}
\end{appendix}

\begin{table}[h]
  \center
  \begin{tabular}{||c|c|c||} \hline
       & no disorder  & RS phase\\ \hline
    $C^z$ & 2   [2.00003(1)]   & 2    \\ \hline
    $C^x$ & 1/2 [0.499998(1)]   & 2    \\ \hline
    $O^z$ & 1/2 [0.499998(1)]   & 0.382 \\ \hline
    $O^x$ & 1/4 [0.249999(2)]   & 0.382 \\ \hline
   \end{tabular}
   \caption {Decay exponents for the XX spin-1/2 chain, values
             obtained by linear fit to data in Fig. \ref{fig01} in brackets.} 
   \label{tab01}
\end{table}

\vfil\eject

\begin{references}
\bibitem{fish1}	Daniel S. Fisher, \prl {\bf 69}, 534 (1992).
\bibitem{fish2} Daniel S. Fisher, \prb {\bf 50}, 3799 (1994).
\bibitem{fish3} Daniel S. Fisher, \prb {\bf 51}, 6411 (1995).
\bibitem{hyma} R.A. Hyman, Kun Yang, R.N. Bhatt and S.M. Girvin, \prl
		{\bf 76}, 839, (1996).
\bibitem{hyma2} R.A. Hyman, Kun Yang, \prl
		{\bf 78}, 1783, (1997).
\bibitem{west1} E. Westerberg, A. Furusaki, M.Sigrist, and
                P. A. Lee, \prl {\bf 75}, 4302, (1995).
\bibitem{west2} E. Westerberg, A. Furusaki, M.Sigrist, and
                P. A. Lee, \prb {\bf 55}, 12578 (1997). 
\bibitem{bale} Leon Balents and Matthew P. A. Fisher, cond-mat/9706069
		(1997).
\bibitem{kenz} Ross H. McKenzie, \prl {\bf 77}, 4804 (1996).
\bibitem{youn} A.P. Young and H. Rieger, \prb {\bf 53}, 8486 (1996).
\bibitem{rode} Heinrich R\"oder, Joachim Stolze, Richard N. Silver and
		Gerhard M\"uller, J. Appl. Phys. {\bf 79}, 4632 (1996).
\bibitem{hida} Kazuo Hida, \prb {\bf 45}, 2207 (1992).
\bibitem{hida2} Kazuo Hida, cond-mat/9707239 (1997).
\bibitem{haas} Stephan Haas, Jose Riera and Elbio Dagotto, \prb
		{\bf 48}, 13174 (1993).
\bibitem{guom} M. Guo, R. N. Bhatt and D. A. Huse, \prl {\bf 72},
               4137 (1994).
\bibitem{rieg} H. Rieger and A. P. Young, \prb {\bf 54}, 3328 (1996).
\bibitem{sent} T. Senthil, cond-mat/9709164 (1997)
\bibitem{hald} F.D.M. Haldane, \pl {\bf 93A}, 464 (1983).
\bibitem{sing} R.R.P Singh {\it et al.}, \prl {\bf 61}, 2484 (1988).
\bibitem{mada} S. K. Ma, C. Dasgupta and C-K Hu, \prl {\bf 43}, 1434
               (1979); C. Dasgupta and S. K. Ma, \prb {\bf 22}, 1305
               (1979)
\bibitem{nijs} M. den Nijs and K. Rommelse, \prb {\bf 40}, 4709 (1989). 
\bibitem{arov} S. M. Girvin and D. P. Arovas, Physica Scripta {\bf T27},
               156 (1989); D. P. Arovas and S. M. Girvin, 
	       ``Exact Questions to Some
	       Interesting Answers in Many Body Physics'',
	       {\em Proc. VIIth Int. Conf. on Recent Progress in 
               Many-Body Theories}, (Plenum Press, New York, 1992), 
               pp. 315-344, editors T.L. Ainsworth, C.E. Campbell, 
               B.E. Clements and E. Kortschek.
\bibitem{kohm} M. Kohmoto and H. Tasaki, \prb {\bf 46}, 3486 (1992).
\bibitem{frad} E. Fradkin, {\em Field Theories of Condensed Matter
		Systems}, (Addison-Wesley Pub. Co., Redwood City, CA., 1991).
\bibitem{lieb} Elliot Lieb, Theodore Schultz and Daniel Mattis,
	       Ann. of Phys. {\bf 16}, 407, 1961.
\bibitem{mcco} Barry M. McCoy, Phys. Rev. {\bf 173}, 531 (1968).
\bibitem{pfeu} Pierre Pfeuty, Ann. of Phys. {\bf 57}, 79 (1969).

\end{references}
\end{document}